\documentclass[3p,times]{elsarticle}

\usepackage{ecrc}

\volume{00}

\firstpage{1}

\journalname{Nuclear Physics A}

\runauth{}

\jid{nupha}

\usepackage{amssymb}

\usepackage[figuresright]{rotating}

\begin{document}

\begin{frontmatter}



\dochead{}

\title{Probing nuclear matter with jets}

 \author[label1,label2]{Ben-Wei Zhang\fnref{label10}}
 \fntext[label10]{Speaker,\,\, {\it Email address}: bwzhang@iopp.ccnu.edu.cn. }
 \author[label1,label2]{Yuncun He}
 \author[label3]{R.~B.~Neufeld}
 \author[label3]{Ivan Vitev}
 \author[label1,label2]{Enke Wang}
 \address[label1]{Institute of Particle Physics, Central China Normal University, Wuhan 430079, China}
 \address[label2]{Key Laboratory of Quark and Lepton Physics (Central China Normal University),
Ministry of Education, China}
\address[label3]{Los Alamos National Laboratory, Theoretical Division, MS B238, Los Alamos, NM 87545, U.S.A.}



\begin{abstract}
Jet physics in relativistic heavy ion collisions, which combines perturbative QCD jet production
with quark and gluon  energy loss and in-medium parton shower modification,  has emerged as a powerful
tool to probe the properties of strongly-interacting matter formed in high-energy nuclear reactions.
We present selected results for the modification of jet cross sections and related observables  in
the ambiance of  hot and/or dense nuclear medium. We focus on the inclusive jet spectrum and dijets
[${\cal O}(\alpha_s^3) $], and  $Z^0/\gamma^*$ tagged jets  [${\cal O}(G_F\alpha_s^2) $]  in the framework
of perturbative QCD.
\end{abstract}

\begin{keyword}

Quark-gluon plasma (QGP) \sep jet production
\sep parton energy loss \sep perturbative QCD.
\end{keyword}

\end{frontmatter}


\section{Introduction}
\label{sec:intro}

Jets, collimated showers of energetic final-state particles,  and the physical observable related to jets
are not only most closely related to the perturbative QCD (pQCD) dynamics, but also  among the
most intensively studied objects in high-energy physics~\cite{Salam:2009jx,EKS}.  Jet productions in elementary
collisions can be accurately calculated and compared to a wealth of experimental
data on jet observables, measured in $e^+ e^-$, deep inelastic scattering and hadron-hadron reactions.
Jets provide not only a laboratory to test and advance pQCD but are also related to important
signatures of new physics (beyond the Standard Model).

QCD theory predicts that in relativistic heavy-ion collisions (HIC) a new kind of matter,
the quark-gluon plasma (QGP) consisting of deconfined quarks and gluons,  may be created.
Theoretical studies and experimental measurements have
shown that fast partons traversing the QGP medium will undergo multiple scattering and
lose energy due to collisional and radiative processes~\cite{GVWZ}.
This parton energy loss and the corresponding suppression of hadron cross sections in nucleus-nucleus (A+A)
reactions have been dubbed  {\it jet quenching}.  Jet quenching effects
 in the hot QCD medium will surely alter the production of reconstructed jets,  which contain one or several
 energetic partons prior to hadronization~\cite{Vitev:2008rz}. The concurrent advance in understanding jet production
and parton energy loss in hot QCD medium has spawned a new direction of research:  the physics of jets
in high-energy nucleus-nucleus collisions at RHIC and at the LHC.

In this manuscript  we discuss  the modification of
jet production in relativistic heavy-ion reactions.  We include selected  results for inclusive jet
productions~ \cite{Vitev:2009rd,He:2011pd}, $Z^0/\gamma^*$-tagged jet
cross sections~\cite{Neufeld:2010fj,Neufeld:2012df} and the dijet transverse momentum
imbalance~\cite{He:2011pd,He:2011sg,HNVZ-2012} in high-energy nuclear collisions.
 Our theoretical simulations will be compared with experimental data on jets when available, and predictions for future measurement are also
presented.

\section{Inclusive jet production in HIC}
\label{sec:dijet}

\begin{figure}[t]
\begin{center}
\includegraphics[width=2.8in,height=2.5in,angle=0]{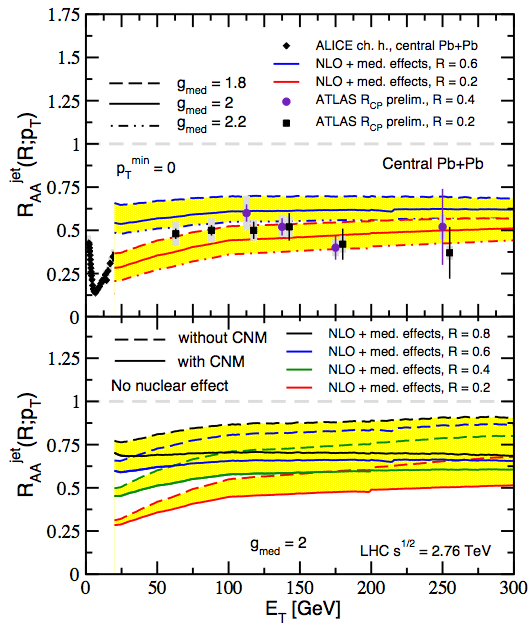}
\hspace{0.6cm}
\includegraphics[width=2.8in,height=2.5in,angle=0]{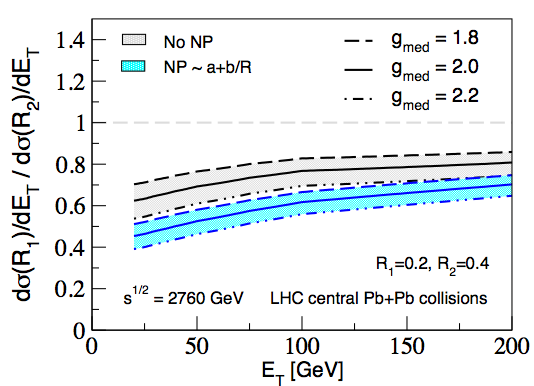} \vspace*{-.1in}
\caption{ Nuclear modification factors for inclusive jets in Pb+Pb
collisions at the LHC (left panel).  Numerical simulation of ratios of inclusive jet cross sections
at different jet size parameter $R$ with and without non-perturbative effect at
 Pb+Pb at  $\sqrt{s_{NN}}=2.76$~TeV  (right panel). }
\label{fig:1-jet}
\end{center}
\end{figure}

Within the pQCD collinear factorization approach, jet cross sections at  $ {\cal O}( \alpha_s^3 ) $
in hadron-hadron collisions can generally be expressed as follows~\cite{EKS,Vitev:2009rd,He:2011pd}:
\begin{eqnarray}
 \!\!\!\! \!\!\! \frac{d\sigma}{dV_{\rm n}}= \frac{1}{2!}\int dV_{\rm 2}
\frac{d\sigma(2\rightarrow2)}{dy_1dy_2d^2E_{T\, 1}d^2E_{T\,2} } \, S_2(p_1^{\mu},p_2^{\mu})
   +\frac{1}{3!}\int dV_{\rm 3} \frac{d\sigma(2\rightarrow3)}
{dy_1dy_2dy_3d^2E_{T\, 1}d^2E_{T\,2} d^2E_{T\,3} }  \,S_3(p_1^{\mu},p_2^{\mu},p_3^{\mu}) \,\, .
   \label{di-pt}
\end{eqnarray}
Here,  $V_{\rm n} = dy_1\cdots dy_nd^2E_{T\,1}\cdots d^2E_{T\,n}$ stands for the multi-parton phase space.
Up to $ {\cal O}( \alpha_s^3 ) $ two
 contributions should be taken into account:   one
from $2 \rightarrow 2$ processes at leading order (LO) and the higher-order virtual corrections  (given by the first term
in above equation). The second one is from $2 \rightarrow 3$ processes, represented by the second term.

In relativistic heavy-ion collisions, energetic partons or jets produced from the hard scattering
will pass through the hot/dense QCD medium and the jet-medium interactions will degrade the
energy of jets and alter the jet shapes.  In our theoretical simulations of jet production in the
nuclear environment we utilize the formalism of in-medium parton splitting~\cite{Ovanesyan:2011kn}, which
in the soft gluon approximation naturally gives the fully differential distribution of the
medium-induced energy loss of fast partons in the QCD medium~\cite{GLV}.

Fig.~\ref{fig:1-jet} illustrates the dependence of the
nuclear modification factor for  single  jet production $R_{AA}^{\rm 1-jet}$
in central Pb+Pb collisions on the jet size $R$ with an acceptance cut without ($p_T^{\min}=0$~GeV)
and with ($p_T^{\min}>0$~GeV)  collisional energy loss~\cite{He:2011pd}.
The left panel  of Fig.~\ref{fig:1-jet} presents $R_{AA}^{\rm 1-jet}$ due to both initial-state and final-state effects
for different jet sizes $R$  and different  coupling constant $g_{\rm  med}$.   The influence of cold nuclear matter effects
is elucidated. The nuclear modification factor $R_{AA}^{\rm 1-jet}$ shows a clear dependence on jet size
because for larger  $R$ more medium-induced gluon radiation will be recaptured within the jet~\cite{Vitev:2009rd}.
However, our results in Fig.~\ref{fig:1-jet} only include radiative energy loss. If collisional energy loss is included,
the dependence of $R_{AA}^{\rm 1-jet}$  on the jet size $R$ will almost disappear~\cite{He:2011pd}.
The right panel  shows the ratio of jet cross sections at two different jet radii at three sets of jet-medium coupling
constant $g_{\rm med}$.  Non-perturbative effects may affect this ratio, as shown in the figure.

\section{Tagged jet production in HIC}
\label{sec:tagged-jet}

\begin{figure}[t]
\begin{center}
\includegraphics[width=2.6in,height=2.2in,angle=0]{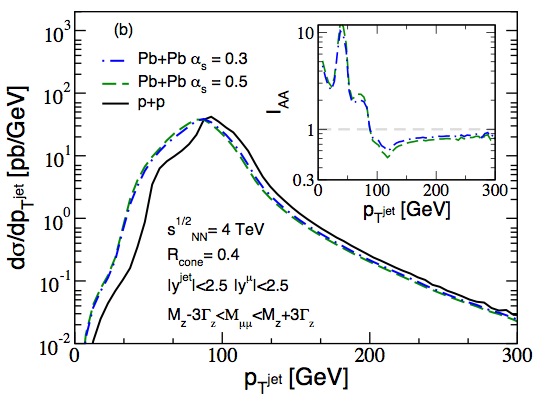}
\hspace{0.6cm}
\includegraphics[width=2.6in,height=2.2in,angle=0]{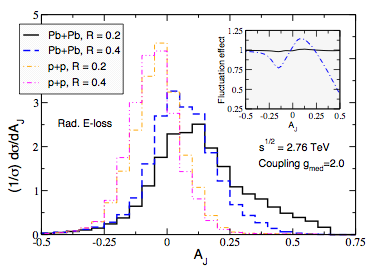} \vspace*{-.1in}
\caption{ Numerical simulations of $Z^0/\gamma^*$ tagged jet transverse momentum spectra in p+p and Pb+Pb
reactions at $\sqrt{s_{NN}}=4.0$~TeV (left panel).
The $Z^0/\gamma^*$ tagged jet transverse momentum asymmetry for p+p and Pb+Pb collisions at
$\sqrt{s} = 2.76$~TeV for two different $R=0.2, 0.4$ (right panel). }
\label{fig:Z-jet}
\end{center}
\end{figure}

To investigate jet production in the nuclear environment and infer the properties of
the QGP, it is important to place  constraints on the magnitude of the parton shower energy
in the medium~\cite{Zhang:2011ak}.  Electroweak vector bosons, such as $Z^0, W^{\pm}$  and $\gamma$,
produced in conjunction with jets can provide such constraints on average~\cite{Neufeld:2010fj,Neufeld:2012df,Dai:2012am}.
For example, in $Z^0$-tagged jet production at  leading-order obeys
the relation \begin{math} p_{T_{\rm jet}} \end{math} = \begin{math} p_{T,Z^0} \end{math}. Next-to-leading order
corrections are very important, as we will see below.
Because $Z^0$  and isolated photons do not interact strongly,
they can escape from the hot medium undisturbed.  By measuring the difference between $p_{T_{\rm jet}}$ and $p_{T_,Z^0}$
in the final state,  one can deduce the average energy loss of the jet in QCD matter.

\begin{figure}[t]
\begin{center}
\includegraphics[width=2.5in,height=1.9in,angle=0]{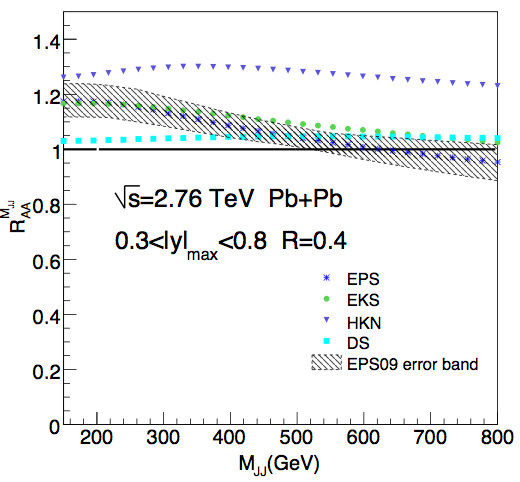}
\hspace{0.6cm}
\includegraphics[width=2.5in,height=1.9in,angle=0]{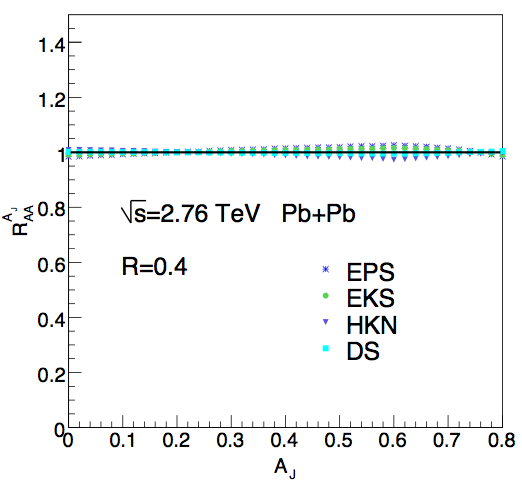} \vspace*{-.1in}
\caption{Nuclear modification factors for the  dijet invariant mass spectra (left panel) and the dijet transverse energy asymmetry distributions (right panel) in Pb+Pb collisions at the LHC with different nPDFs. }
\label{fig:di-1}
\end{center}
\end{figure}

Recently the first  ${\cal O}(G_F\alpha_s^2)$\ calculations on $Z^0/\gamma^*$-tagged jet production
and event asymmetry in HIC with final-state
parton energy loss effect have been done~\cite{Neufeld:2010fj,Neufeld:2012df}.  This approach
has been extended to study isolated-photon tagged jets in A+A collisions
at  ${\cal O}(\alpha_{\rm em} \alpha_s^2)$~\cite{Dai:2012am}. The left panel of Fig.~\ref{fig:Z-jet} shows
$Z^0/\gamma^*$ tagged jet transverse momentum spectra in p+p and Pb+Pb at
$\sqrt{s} = 4,0$~TeV. Due to parton energy loss effects, the tagged jet spectra are shifted
to smaller $ p_{T_{jet}}$ values.
 If we plot the ratio of  the cross section in Pb+Pb reaction (per binary collision) to the one in p+p, as shown in the
 insert of the left panel figure,
we observe a transition from the enhancement at small momentum to suppression at the large momemtum
~\cite{Neufeld:2010fj}.
The right panel  of Fig.~\ref{fig:Z-jet}   presents the event asymmetry, where we define
  $A_J = ({p_T}_1 - {p_T}_2)/({p_T}_1 + {p_T}_2) $
whith ${p_T}_1$ and ${p_T}_2$ being the transverse momenta of the leading and subleading jets. For the case of
$Z^0/\gamma^*$ tagged jet production ${p_T}_1=p_{T,Z^0}$ and  ${p_T}_2=p_{T_{\rm jet}} $.  We observe that
with jet-medium interaction the tagged jet $A_J$ distributions in Pb+Pb reactions are significantly  broader
and shifted to   $A_J > 0$. The underlying physics mechanism is medium-induced parton splitting:
energy is lost due to large-angle radiation out of the jet reconstruction parameter
$R$.  Consequently, the smaller  the jet radius, the larger
the width and the average asymmetry of these distributions will be~\cite{Neufeld:2012df} .

\section{Dijet asymmetry in high-energy nucleus-nucleus collisions}
\label{sec:dijet}

Recently, the ATLAS and CMS collaborations at the LHC~\cite{Aad:2010bu,Chatrchyan:2011sx} published
results on the enhancement of the transverse energy imbalance
for dijet in Pb+Pb collisions.  This was the first evidence for the quenching of reconstructed jets at the LHC.
In the framework of pQCD,  we investigate dijet productions to $ {\cal O}( \alpha_s^3 ) $ at RHIC and the LHC
 by including both initial- and final-state nuclear matter effects~\cite{He:2011pd,He:2011sg}.

The dijet invariant mass $M_{JJ}$ is defined as the invariant mass of all particles in the final state, $[(\sum p^{\mu}_n)^2]^{1/2}$.  Cold nuclear matter(CNM) effects on $M_{JJ}$ and $A_J$ distributions in Pb+Pb collisions at $\sqrt{s}=2.76$~TeV~ are
displayed in Fig.~\ref{fig:di-1}  by implementing four parametrization sets of nuclear parton distributions (nPDFs). The left
panel shows dijet invariant mass spectra are enhanced in a wide region of $M_{JJ}$ due to CNM effects, which is opposite to the suppression resulting from the final-state QGP effects. The right panel  of  Fig.~\ref{fig:di-1}  shows that the dijet asymmetry
distribution is insensitive to the CNM effects and, thus, provides a robust observable to probe the final-state effects in the QGP.

\begin{figure}[t]
\begin{center}
\includegraphics[width=2.5in,height=2.0in,angle=0]{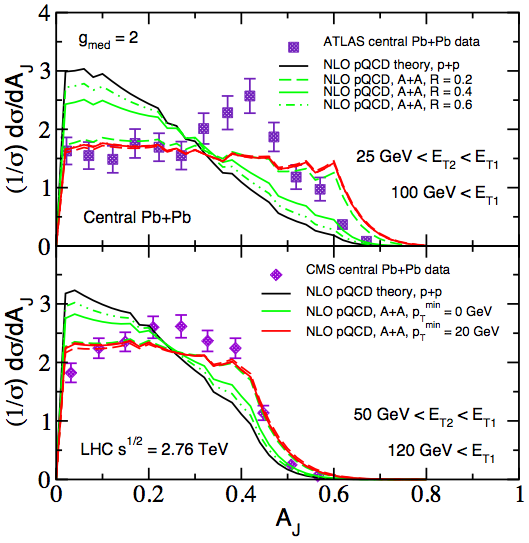}
\hspace{0.6cm}
\includegraphics[width=2.5in,height=2.0in,angle=0]{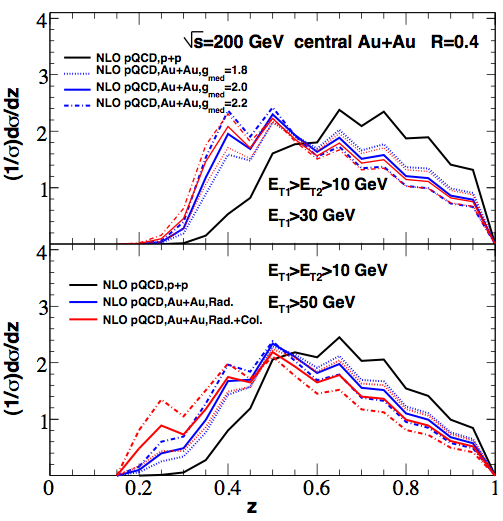} \vspace*{-.1in}
\caption{ The dependence of the normalized dijet asymmetry distributions  $A_J$ on
the jet radii
 in Pb+Pb at the LHC (left panel). Similar dependencies on coupling strength are studied and for momentum imbalance
 $z$ distribution in Au+Au reactions at  RHIC (right panel).  }
\label{fig:di-2}
\end{center}
\end{figure}

The energy fraction $z=E_{T2}/E_{T1}$, which represents the transverse energy imbalance of dijet  production,
plays a similar role to  the asymmetry  $A_J$. The normalized dijet  asymmetry
distributions in central Pb+Pb collisions at $\sqrt{s}=2.76$~TeV and  imbalance distribution in central Au+Au
collisions at $\sqrt{s}=200$~GeV are presented in Fig.~\ref{fig:di-2}. We  see that  in A+A collisions
these distributions are enhanced due to the energy loss of jets propagating the QGP.
For radiative energy losses, there is distinct dependence on the jet cone radius R.
When  the collisional energy dissipation of the art on shower is also considered, the dependence of
imbalance distributions on jet sizes is very weak,
but the sensitivity to the coupling strength of jet to medium is amplified~\cite{He:2011pd}. Furthermore,
jet-medium interactions  shift the $z$ distribution to the left and,
thus, give smaller averaged $\langle z\rangle $ in Au+Au collisions~\cite{HNVZ-2012}, see the right
panel of Fig.~\ref{fig:di-2}.

\end{document}